# Accretion disc turbulence and the X-ray power spectra of black hole high states

Michael A. Nowak [1,2] and Robert V. Wagoner [3]

[1] CITA, 60 St. George St., Toronto, Ontario M5S 1A7
[2] Current address: JILA, Campus Box 440, Boulder, CO 80309-0440
[3] Department of Physics, Stanford University, Stanford, CA 94305-4060

**ABSTRACT**

The high state of black hole candidates is characterized by a quasi-thermal emission component at $kT \sim 1$ keV. In addition, this state tends to have very low variability which indicates that it is relatively stable, at least on *short* time scales. Most models of the high state imply that the bulk of the emission comes from an optically thick accretion disc; therefore, this state may be an excellent laboratory for testing our ideas about the physics of accretion discs. In this work we consider the implications of assuming that accretion disc viscosity arises from some form of turbulence. Specifically, we consider the simple case of three dimensional hydrodynamic turbulence. It is found that the coupling of such turbulence to acoustic modes in the disc can alter the disc emission. We calculate the amplitude and frequencies of this modulation, and we express our results in terms of the X-ray power spectral density. We compare our calculations with observations of the black hole candidate GS 1124-683, and show that for certain parameters we can reproduce some of the high frequency power. We then briefly explore mechanisms for producing the low frequency power, and note the difficulty that a single variability mechanism has in reproducing the full range of observed variability. In addition, we outline ways in which future spacecraft missions – such as USA and XTE – can further constrain our model, especially at frequencies above $\sim 10^2$ Hz.

## 1 INTRODUCTION

Black hole candidates (BHC) are typically identified either from estimates of their mass, via a binary mass function, or by analogies with other BHC where one typically compares high energy emission and variability properties. (For a review of black hole candidates, including a summary of observed spectral and temporal properties, see Tanaka & Lewin 1994.) The energy spectra of BHC have been historically labelled based upon observations of the soft X-ray band ($\sim 1-10$ keV). Intense, quasi-thermal flux is referred to as the "high" state. Non-thermal flux (typically a power law with a photon index of $\sim 1.7$) in this band indicates that the BHC is in the "off" state, for extremely low intensity flux, or "low" state, for moderate intensity flux that *in the $\sim 1-10$ keV band* is less than the high state flux in the same band. [It is possible, however, for the *bolometric* luminosity to be greater in the low state than in the high state. See the comments by Tanaka & Lewin (1994) and Nowak (1994b).] The high state tends to have little variability, with a root mean square (RMS) variability of a few percent, whereas the low state tends to have an RMS variability of several tens of percent. (We will make explicit what we mean by RMS variability in §4.)

Many models have been proposed to explain the observed spectra. Ultimately, nearly all of them involve the compact object accreting matter via an accretion disc. For the case of the non-thermal low state, this disc has been modelled as being optically thin (*cf.* Melia & Mesra 1993,

Luo & Liang 1994), as a cold reflector of a central power law source (*cf.* Done et al. 1992), and as shrouded in a hot, optically thick, Comptonizing cloud (*cf.* Sunyaev & Titarchuk 1980). On the other hand, the most commonly accepted model for the quasi-thermal high state is optically thick emission from a relatively stable and quiescent disc (*cf.* Shakura & Sunyaev 1973). Though the details of the high state solutions differ, the one common feature that they have is the emission domination of the disc itself. That is, there is no need to invoke clouds, winds or other external sources of emission. Recent models have been very succesful at explaining luminosity variations in this state as arising from accretion rate variations in a disc of constant inner radius, lending further support to the accretion disc picture (*cf.* Ebisawa et al. 1993, Ebisawa 1994). In addition, the fact that the RMS variability of the high state is very low suggests that the disc is in a relatively steady state, at least on short time scales. These properties of the high state – predominantly disc-dominated and steady accretion – are what make it the ideal laboratory for testing our notions about the physics of accretion discs.

The purpose of this work is to explore how one aspect of accretion disc physics, namely disc turbulence, may manifest itself in the observed properties of BHC high states. Specifically, we will explore the ways in which turbulence affects the observed variability of the high state. In the next section, we will outline our basic assumptions about the nature of accretion disc turbulence. For our purposes, we shall only consider isotropic hydrodynamic turbulence. (We will explore the more realistic cases of anisotropic and/or magnetohydrodynamic turbulence in future works.) In §3 we will consider two ways in which turbulence may affect the observed disc emission, namely, advection of photons from the hot disc midplane and direct modulation of the emission rate. The latter will be seen to be the more important of these two. This manifestation of turbulence is essentially the coupling of acoustic modes to the turbulent eddies. In §4 we consider how the emission modulation is revealed in the observed X-ray power spectral density of BHC high states. We compare our predictions to high state observations of GS 1124-683 that cover the frequecy range $10^{-2} - 10^2$ Hz. We will see that for certain parameter ranges, acoustic mode coupling to turbulence can account for the high frequency power. In §5 we briefly explore whether weak fluctuations on viscous or thermal time scales can explain the observed low frequency power. Finally, in §6 we summarize our results and discuss both future theoretical and observational prospects.

## 2 BASICS OF THE DISC TURBULENCE

The single greatest uncertainty in modelling accretion discs is the viscosity prescription. This uncertainty directly affects our understanding of mass and angular momentum transport in the disc, but more importantly for our purposes it affects our understanding of the energy generation (and hence structure) within the disc. Most models require a phenomenological description of the radial energy generation profile and disc structure; however, here we require a model of this structure in the vertical direction (*i.e.*, perpendicular to the disc midplane) as well. In this work we shall explore the consequences of assuming that viscosity is ultimately due to accretion disc turbulence.

One of the original motivations for the $\alpha$-disc model of Shakura & Sunyaev (1973) was consideration of (locally) homogeneous and isotropic hydrodynamic turbulence. Applications and modifications of the $\alpha$-disc model have grown beyond this initial motivation; however, in this work we shall return to the viewpoint that viscosity is directly related to the properties of turbulent eddies. Even within this framework, recent works have considered models more sophisticated than homogeneous and isotropic hydrodynamic eddies. Some researchers have carefully considered the role of rotation and have allowed for anisotropic (*i.e.* two-dimensional) eddies (*cf.* Narayan et al. 1994, Kato & Yoshizawa 1993, Dubrulle 1992). One of the more exciting developments has been the study of the magnetic shearing instability (MSI), which leads to three-dimensional magneto-hydrodynamic turbulence (Balbus & Hawley 1991, Hawley & Balbus 1991, Balbus et al. 1994). Similar to classical isotropic hydrodynamic turbulence, the MSI turbulence is quasi-incompressible. However, its relation to energy generation in a disc, and its coupling to the global dynamics of the accretion disc, are as of yet unknown. In this work we will only consider isotropic, hydrodynamic turbulence as embodied in the $\alpha$-disc model. We choose this model as being the most analytically tractable and as being able to provide a guide for using more sophisticated models (namely anisotropic eddies and MSI turbulence) in future work. For the basic physics of this model we will follow the work of Shakura & Sunyaev (1973) and the discussion of turbulence found in Landau & Lifshitz (1987).

At a given radius in the disc, let the largest turbulent eddy size be $L$, and have an eddy velocity $u_L$, such that

$$u_L \sim \mathcal{M}_L c_s \quad, \quad L \sim \delta h \quad, \tag{2.1}$$

where $\mathcal{M}_L$ is the Mach number of the turbulence (assumed $\lesssim 1$), $c_s \sim h\Omega$ is the disc speed of sound, $h$ is the half-thickness of the disc, $\Omega$ is the Keplerian rotation velocity, and $\delta$ is a dimensionless scaling factor. In the vertical direction the eddy size is limited by the disc thickness. In the radial direction, constraining the disc velocity difference across the eddy to being less than the speed of sound also limits the eddy size to $\lesssim h$. The turbulent viscosity is found by taking the simplest combination of the above quantities that has the dimensions of a kinematic viscosity. We therefore have for the turbulent viscosity

$$\nu_t \sim u_L L \sim \mathcal{M}_L \delta h^2 \Omega \sim \alpha h^2 \Omega \quad. \tag{2.2}$$

In the $\alpha$-disc model the viscous azimuthal stress is given by $f_\phi \sim \nu_t \rho \Omega \sim \alpha P$, where $P \sim \rho c_s^2$ is the pressure. This gives us the above identification $\alpha \sim \mathcal{M}_L \delta$.

Assuming that the standard form for the viscosity tensor holds for our Keplerian disc, the rate of viscous energy dissipation per unit mass is given by

$$\dot{\epsilon} \sim \nu_t \left( r \frac{\partial \Omega}{\partial r} \right)^2 \sim u_L L \Omega^2 \quad. \tag{2.3}$$

However, from the standard Kolmogorov turbulence scaling laws (*cf.* Landau & Lifshitz 1987) we have

$$u_L \sim (\dot{\epsilon} L)^{1/3} \sim (u_L L^2 \Omega^2)^{1/3} \implies u_L \sim L\Omega \implies \delta \sim \mathcal{M}_L \quad. \tag{2.4}$$

We can then identify

$$\alpha \sim \mathcal{M}_L^2 \quad. \tag{2.5}$$

Since $\alpha$ is a direct measure of the Mach number of the turbulence, it is directly related to the variations in the energy generation rate, as we will show in subsequent sections.

Using $\alpha$ as a measure of the turbulent Mach number of the largest eddies, we find the turbulent velocity, $u_L$, eddy size, $L$, and eddy lifetime or rollover time, $t_L$, of the largest scale eddies to be given by

$$u_L \sim \sqrt{\alpha}h\Omega \quad, \quad L \sim \sqrt{\alpha}h \quad, \quad t_L \sim L/u_L \sim \Omega^{-1} \quad . \tag{2.6}$$

On smaller scales, $\ell$, we use the standard Kolmogorov scaling laws to find

$$u_\ell \sim \alpha^{1/3} h\Omega \left(\frac{\ell}{h}\right)^{1/3} \quad, \quad t_\ell \sim \ell/u_\ell \sim \alpha^{-1/3}\Omega^{-1}\left(\frac{\ell}{h}\right)^{2/3} \quad, \tag{2.7}$$

where $\ell \lesssim \sqrt{\alpha}h$. Note that for the smaller scales, the turbulent Mach number $\mathcal{M}_\ell \sim \mathcal{M}_L^{2/3}(\ell/h)^{1/3}$.

The pressure variation associated with the largest eddy is given by

$$\Delta p_L \sim \rho u_L^2 \sim \alpha \rho h^2 \Omega^2 \quad, \tag{2.8}$$

and therefore on smaller scales, $\ell$, we have

$$\Delta p_\ell \sim \rho u_\ell^2 \sim \alpha^{2/3}\rho h^2 \Omega^2 \left(\frac{\ell}{h}\right)^{2/3} \quad . \tag{2.9}$$

We have used the scaling laws for incompressible turbulence; however, we expect weak density fluctuations to be associated with these eddies. The density fluctuations should be given by $\Delta \rho \sim c_s^{-2}\Delta p$, therefore we have

$$\Delta \rho_L \sim \mathcal{M}_L^2 \rho \quad, \quad \Delta \rho_\ell \sim \mathcal{M}_\ell^2 \rho \quad . \tag{2.10}$$

We can consider these density fluctuations as being due to acoustic modes coupled to the turbulence (*cf.* Goldreich & Kumar 1988). Essentially, eddies of frequency $f_\ell \equiv t_\ell^{-1}$ will couple to acoustic modes of the same frequency or lower. In this work, we shall calculate the coupling to azimuthally symmetric ($m = 0$) modes, as they are observationally most relevant. [Modes with multiple nodes in the azimuthal direction ($m \geq 1$) will have the azimuthally integrated flux reduced to some degree.] We shall use equation (2.10) as an estimate of the amplitude of this coupling for the calculations of §4.

## 3  INFLUENCE OF TURBULENCE ON THE DISC SPECTRUM

In this section we shall consider ways in which the accretion disc turbulence might manifest itself in the observed X-ray variability of the source. We consider two possible processes: advection and direct modulation of emission. If the typical photon radiative diffusion time from the disc midplane is much longer than a typical eddy rollover time, there exists the possibility that eddies will advect photons from the hot disc midplane to the cooler disc surface. (For a Shakura-Sunyaev $\alpha$-disc model, the midplane temperature is typically a factor of $\sim 10$ greater than the disc surface temperature, though this depends upon such parameters as $\alpha$ and the accretion rate; Shakura & Sunyaev 1973, Shapiro & Teukolsky 1983.) One must also check to see that the photon absorption time is longer than the eddy rollover time, else the photon will be absorbed and reemitted with properties of the local temperature.

For the case of emission modulation, if one assumes that the energy generation rate per unit volume scales as the local density, then the emission can be modulated by the weak density fluctuations associated with the turbulence. In this case, the photon diffusion time (though not necessarily the photon absorption time) must be longer than the eddy lifetime in order for the over/under-density to be manifested in emission variability. We shall explore each of these possibilities separately below.

*a) Advection*

In order for advection to be effective at "dredging" up photons from the hot midplane of the disc, the photons must undergo many scatterings during the eddy rollover time. We assume that these scatterings will occur as the result of electron scattering opacity, $\kappa_{es} = 0.4$ cm$^2$ g$^{-1}$. The photon diffusion time due to scattering across an eddy length, $\ell$, must be much greater than the rollover time of the eddy, $t_\ell$. (Equivalently, the photon diffusion length over a time $t_\ell$ must be much less than $\ell$.) We shall ignore the effects of absorption in the following, which would serve to make our constraints even more severe.

The number of scatterings, $N_\ell$, over a diffusion length, $\ell$, is that of a random walk (*cf.* Rybicki & Lightman 1979), namely

$$N_\ell^{1/2} \sim \ell/\lambda_{es} \sim \kappa_{es}\rho\ell \equiv \tau_\ell^{es} \ , \tag{3.1}$$

where $\lambda_{es}$ is the electron scattering mean free path and $\tau_\ell^{es}$ is defined to be the electron scattering optical depth over length $\ell$. The total path length travelled by the photon is given by

$$N_\ell \lambda_{es} \sim N_\ell^{1/2} \ell \sim \tau_\ell^{es} \ell \equiv \tau_h^{es} \, h \left(\frac{\ell}{h}\right)^2 \ , \tag{3.2}$$

where $\tau_h^{es}$ is the electron scattering optical depth over the disc half-thickness $h$. The diffusion time over length $\ell$ is therefore

$$t_\ell^D \sim \tau_h^{es} \frac{h}{c} \left(\frac{\ell}{h}\right)^2 \ , \tag{3.3}$$

where $c$ is the speed of light. For advective transport to be significant, this diffusion time must be much greater than the eddy rollover time:

$$\tau_h^{es} \frac{h}{c} \left(\frac{\ell}{h}\right)^2 \gg \alpha^{-1/3} \Omega^{-1} \left(\frac{\ell}{h}\right)^{2/3} \ . \tag{3.4}$$

Accounting for the maximum eddy size as given in equation (2.6), the above implies that

$$\sqrt{\alpha} \gtrsim \left(\frac{\ell}{h}\right) \gg \alpha^{-1/4} \left(\tau_h^{es}\right)^{-3/4} \left(\frac{c_s}{c}\right)^{-3/4} \ . \tag{3.5}$$

In a previous work (Nowak & Wagoner 1992) we have noted that for a Shakura-Sunyaev $\alpha$-disc, $c_s/c \sim \beta^{-1}\mathcal{L}$, and that $\Sigma \equiv 2\rho h \sim \beta^2 (\alpha \, \kappa_{es} \, \mathcal{L})^{-1}$, where $\beta$ is the ratio of radiation pressure to total pressure and $\mathcal{L}$ is the ratio of the disc luminosity to the Eddington luminosity of the central compact object. Using these estimates, we find that $\tau_h^{es} \, c_s/c \sim \beta/\alpha$, and therefore for photon advection to be effective in advecting photons from the midplane to the surface (giving the greatest temperature contrast) we require

$$\sqrt{\alpha} \gtrsim \left(\frac{\ell}{h}\right) \gg \beta^{-3/4}\sqrt{\alpha} \ . \tag{3.6}$$

Since $\beta \lesssim 1$, this effectively excludes photon advection as an important process for modulating the disc output.

*b) Emission*

The unimportance of advection can also be an advantage. The fact that a photon will escape the turbulent eddy in which it was created in less than an eddy lifetime means that photons will retain knowledge of the eddy's over/under-density. That is, modulation of emission will not be smeared out by diffusion over many eddy lifetimes. The problem of emission modulation breaks up into two parts – the atmosphere above effective optical depth $\tau^* \sim \sqrt{\tau^{abs}(\tau^{abs} + \tau^{es})} \sim 1$ (which produces an optically thin spectrum and has the potential to effect the shape of the emergent spectrum via electron scattering); and the atmosphere below an effective optical depth $\tau^* \sim 1$ (where the bulk of the energy generation occurs). For both parts of this problem, modulation from the small scale (high frequency) eddies at large depths will be smeared out because photons diffuse through the remaining atmosphere above the eddy (typically a greater length than the eddy size itself) in times longer than the eddy lifetime.

First let us consider emission from depth $\tau^* \lesssim 1$. The diffusion time from a vertical depth, $z$, is given by

$$t_z^D \sim \tau_h^{es} \frac{h}{c} \left(\frac{z}{h}\right)^2 . \tag{3.7}$$

In order that any modulation not be suppressed, we require $t_z^D \lesssim t_\ell$ (*cf.* equation[3.4]). This implies that for modulation due to an eddy of size $\ell$ we are restricted to depths of

$$\left(\frac{z}{h}\right) \lesssim \alpha^{-1/6} \left(\tau_h^{es} \frac{c_s}{c}\right)^{-1/2} \left(\frac{\ell}{h}\right)^{1/3} . \tag{3.8}$$

With our previous estimates of the disc parameters, this becomes

$$\left(\frac{z}{h}\right) \lesssim \alpha^{1/3} \beta^{-1/2} \left(\frac{\ell}{h}\right)^{1/3} . \tag{3.9}$$

(Note that the above depth is always greater than an eddy size, in agreement with the results of the previous section.) As an example, for typical accretion disc parameters ($\alpha \sim 0.1$, $M \sim M_\odot$) effective optical depth $\tau^* \sim 1$ occurs at $z/h \sim 0.01 - 0.02$, which implies that in order for an eddy to modulate the emission, it must have

$$\frac{\ell}{h} \gtrsim 10^{-4} - 10^{-5} . \tag{3.10}$$

This covers a fairly broad range of eddy sizes and frequencies; however, being that this is an effective optical depth $\tau^* \lesssim 1$, the modulation will not affect the bulk of the emitted energy which mainly originates at optical depths $\tau^* \gtrsim 1$.

We now consider the more important mechanism of diffusion from the energy generating part of the atmosphere at $\tau^* > 1$. The diffusion time outward *from the midplane of the disc* is given by:

$$\tau_h^* t_{max}^D \sim \tau_h^* \tau_h^{es} \frac{h}{c} \left(\frac{\Delta z_{max}}{h}\right)^2 \sim \frac{\tau_h^{es}}{\tau_h^*} \frac{h}{c} , \tag{3.11}$$

where $t_{max}^D$ is the diffusion time over a distance $\Delta z_{max}$ equal to the thickness of effective optical depth unity, and $\tau_h^*$ is the total effective optical depth of the atmosphere. Implicit in the above is the condition $1 < \tau_h^* < \tau_h^{es}$. We require the diffusion time from the midplane to be less than an eddy lifetime, implying

$$\alpha^{-1/3}\,\Omega^{-1}\left(\frac{\ell}{h}\right)^{2/3} \gtrsim \frac{\tau_h^{es}}{\tau_h^*}\frac{h}{c}\ , \tag{3.12}$$

which yields

$$\frac{\ell}{h} \gtrsim \sqrt{\alpha}\left(\frac{\tau_h^{es}}{\tau_h^*}\frac{c_s}{c}\right)^{3/2}\ . \tag{3.13}$$

Eddies of this size are able to modulate the emission on their rollover time scale *even at the greatest depths*.

We can evaluate the above inequality for the X-ray producing regions of Shakura-Sunyaev $\alpha$-discs. These discs can be broken up into either radiation pressure dominated regions or gas pressure dominated regions. Utilizing the disc solutions of Nowak (1992) (based upon those of Shakura & Sunyaev 1973), the above inequality becomes

$$\frac{\ell}{h} \gtrsim \sqrt{\alpha}\,(1-1000)\,\mathcal{L}^3\ , \tag{3.14}$$

in the radiation pressure dominated regime, and

$$\frac{\ell}{h} \gtrsim \sqrt{\alpha}\,(10^{-3}-5)\,\mathcal{L}^{21/20}\ , \tag{3.15}$$

in the gas pressure dominated regime. That is, as long as $\mathcal{L} \lesssim 0.1$, *at least* the largest eddies (with $\ell/h \sim \sqrt{\alpha}$) will be able to modulate the emission. Over a wide range of radii and shallower depths, smaller (and hence higher frequency) eddies will also be able to modulate the emission of an accretion disc. In the next section, we explore the amplitude and form of this modulation for accretion discs about galactic black hole candidates.

## 4 RELATIONSHIP TO THE POWER SPECTRAL DENSITY

We have seen above that accretion disc turbulence will mainly affect the observed flux by coupling to acoustic modes which in turn alter the emission from various radii in the disc. The question now becomes one of how these acoustic modes manifest themselves in the observed variability as characterized by the normalized X-ray power spectral density (PSD). Given an observed photon number count rate, $x$, the PSD is formed from the square of the Fourier transform of $x$. Integrating the PSD over all Fourier frequencies yields a quantity that is proportional to $\langle x^2 \rangle$, where the brackets indicate an average over the photon count rate time-series. For the particular PSD normalization that we shall use, integrating the PSD over $f > 0$ yields: $(\langle x^2 \rangle - \langle x \rangle^2)/\langle x \rangle^2$ (the square root of this quantity is the RMS variability, *cf.* Miyamoto et al. 1992). Note that with this normalization the PSD is independent, to within noise, of source luminosity and distance.

We assume that the photon intensity from the disc is directly proportional to the disc's local energy generation rate. That is, we shall ignore variations in the shape of the emergent photon spectra. (This assumption is equivalent to saying that locally $\Delta \ln L \gg \Delta \ln T$, that is the fractional change in the luminosity from a given radius is greater than the fractional change in the temperature.

For a blackbody spectrum, this assumption is essentially the same as saying $\Delta \ln T^4 \sim \Delta \ln T^3$, which we consider to be accurate enough for our crude calculations.) Furthermore, we assume that the disc's energy generation rate scales as the local density to a power of $\mathcal{O}(1)$. This assumption is implicit in many $\alpha$-disc models (*cf.* Shakura & Sunyaev 1973), though some models assume that the energy generation rate is proportional to the local pressure (*cf.* Laor & Netzer 1989). The two assumptions are equivalent in the gas pressure dominated regions of the disc, but not in the radiation pressure dominated regions. For our purposes, we take the variations in the photon flux to be directly proportional to the variations in the density.

The mean square photon count rate, $\langle x^2 \rangle$, is composed of two parts: the mean (DC) intensity squared, plus an incoherent sum of oscillating acoustic sources. Thus we have:

$$\begin{aligned} \langle x^2 \rangle &= \langle x \rangle^2 + \sum_f \begin{pmatrix} \#\ of\ modes\ at \\ frequency\ f \end{pmatrix} \langle \Delta x_f^2 \rangle \\ &= \langle x \rangle^2 + \int_V dV \int_f df \begin{pmatrix} \#\ of\ modes\ per \\ frequency\ per\ volume \end{pmatrix} \langle \Delta x_f^2 \rangle \ , \end{aligned} \quad (4.1)$$

where $\langle \Delta x_f^2 \rangle$ is the mean square photon count rate due to a single mode and where the integrations are over the volume of the accretion disc and over the frequencies of the acoustic modes. For each individual mode we can write

$$\langle \Delta x_f^2 \rangle = \left\langle \left[ C \times \Delta \rho \times \begin{pmatrix} Mode \\ Volume \end{pmatrix} \right]^2 \right\rangle \ , \quad (4.2)$$

where $C\rho = $ photons/sec/volume, or $C\rho \Delta V \sim \Delta x$. The constant, $C$, is related to the disc photon flux $\mathcal{F}$ via

$$C \rho h \sim \frac{dx}{2\pi r\ dr} \sim \mathcal{F}(r) \ , \quad (4.3)$$

where $r$ is the radial coordinate of the disc and $h$ is the disc half-thickness. We therefore have for the mean square photon count rate

$$\langle x^2 \rangle = \langle x \rangle^2 + \int_V dV \int_f df \left( \frac{d^2 N}{dV df} \right) \frac{\mathcal{F}^2(r)}{h^2} v^2(f) \left\langle \left( \frac{\Delta \rho}{\rho} \right)^2 \right\rangle \ , \quad (4.4)$$

where $f^{-2}\ d^2 N/dV df$ is the acoustic mode phase space density, and $v(f)$ is the acoustic mode volume. Note the similarity of the above to the previous definition of the integrated PSD. We therefore identify as the normalized PSD

$$\mathcal{P}(f) = \frac{\int dr\ 2\pi r \left( \frac{d^2 N}{dV df} \right) \frac{\mathcal{F}^2(r)}{h} v^2(f) \left\langle \left( \frac{\Delta \rho}{\rho} \right)^2 \right\rangle}{\left( \int dr\ 2\pi r\ \mathcal{F}(r) \right)^2} \ . \quad (4.5)$$

In the above definition of the normalized PSD, the unperturbed accretion disc model provides $\mathcal{F}(r)$ and $h$, the model of turbulence provides $\Delta \rho / \rho$, and the nature of the acoustic modes provides $d^2 N/dV df$ and $v(f)$. We shall only study the case of axially symmetric modes as these will be be the most readily observable (being of uniform brightness around a ring, there are no cancellations in the observed flux modulation). The dispersion relation of acoustic modes in a rotating disc is given by

$$\omega^2 \approx c_s^2 k^2 + \kappa^2 \quad , \tag{4.6}$$

where $\omega$ is the mode frequency, $c_s$ is the speed of sound, $k$ is the mode wave number, and $\kappa$ is the epicyclic frequency ($\kappa^2 \equiv 4\Omega^2 + r\partial\Omega^2/\partial r$, where $\Omega$ is the Keplerian frequency, cf. Nowak & Wagoner 1991). In terms of the mode frequency, $f$, given in Hz, this can be written as

$$f \approx \sqrt{\frac{\kappa^2}{(2\pi)^2} + c_s^2 \left(\lambda_r^{-2} + \lambda_z^{-2}\right)} \quad . \tag{4.7}$$

With this dispersion relation, the mode volume becomes

$$v \sim 2\pi r \lambda_r \lambda_z \sim 2\pi r \left(\frac{c_s}{f_L}\right)^2 \left(\frac{f_L}{f}\right)^2 \sim 2\pi r h^2 \left(\frac{f}{f_i}\right)^{-2} \left(\frac{f_L}{f_i}\right)^2 \quad . \tag{4.8}$$

In the above, $f_L = \Omega$ is the *local* Keplerian frequency, i.e. the frequency of the *largest* turbulent eddies, and $f_i$ is the rotation frequency, in Hz, of the disc inner edge. Taking the frequency derivative of the inverse of the mode volume, we find the phase space density to be

$$\begin{aligned}\frac{d^2 N}{dV df} &\approx \frac{d}{df} (2\pi r \lambda_r \lambda_z)^{-1} = \frac{1}{2\pi r}\left(\frac{\lambda_r}{\lambda_z} + \frac{\lambda_z}{\lambda_r}\right)\frac{f}{c_s^2} \\ &\sim (\pi h^2 r)^{-1}\frac{1}{f_i}\left(\frac{f}{f_i}\right)\left(\frac{f_L}{f_i}\right)^{-2} \quad . \end{aligned} \tag{4.9}$$

(Note that in equations [4.8] and [4.9] we have taken a "worst case scenario" of $\lambda_r \sim \lambda_z \sim h$.)

For the model of acoustic turbulence that we have chosen, the density fluctuation, and hence the local photon count rate fluctuation, depends upon

$$\frac{\Delta \rho_\ell}{\rho} \sim \mathcal{M}_\ell^2 \sim \mathcal{M}_L^{4/3}\left(\frac{\ell}{h}\right)^{2/3} \sim \frac{f_L}{f_\ell} \quad . \tag{4.10}$$

Thus as a function of frequency, $f$, we have

$$\left(\frac{\Delta \rho}{\rho}\right)^2 \sim \alpha^2 \left(\frac{f}{f_i}\right)^{-2} \left(\frac{f_L}{f_i}\right)^2 \quad . \tag{4.11}$$

Below we briefly will explore other frequency dependences for the density fluctuation, but we will see that it makes no difference in the observable part of the PSD.

Note that in equations (4.8-4.11) we wrote the quantities in terms of $f_L/f_i$, the local Keplerian frequency divided by the Keplerian frequency of the disc inner edge. We do this because the one-to-one correspondance between disc radius, $r$, and Keplerian frequency, $f_L$, allows us to replace the radial integration in equation (4.5) with a frequency integration. Performing a change of variables, we replace $r$ and $dr$ with

$$r \sim 6\left(\frac{f_L}{f_i}\right)^{-2/3} \quad \text{and} \quad dr \sim -4\left(\frac{f_L}{f_i}\right)^{-5/3} d\left(\frac{f_L}{f_i}\right) \quad . \tag{4.12}$$

We must keep in mind that at each radius, the local Keplerian frequency is the *lowest* turbulent frequency. That is, locally turbulence will excite acoustic modes with frequencies greater than or equal to the Keplerian frequency. If we consider the PSD at a given frequency, $f$, it will be made up of acoustic modes that range over radii from $r \to \infty$ to a radius $r$ such that $\Omega(r) = f$ (for $f < f_i$, otherwise $r$ cuts-off at the disc inner edge). We can therefore write for the normalized PSD

$$\mathcal{P}(f) \sim 3 \times 10^{-2} \, \mathcal{N}^{-1} \, \frac{\alpha^2}{f_i} \left(\frac{f}{f_i}\right)^{-5} \int_0^{\min(f/f_i,1)} h \mathcal{F}^2 \left(\frac{f_L}{f_i}\right) d\left(\frac{f_L}{f_i}\right) \, . \qquad (4.13)$$

In the above, $\mathcal{N}$ is the normalization integral

$$\mathcal{N} \equiv \left(\int_6^\infty dr \, r \, \mathcal{F}(r)\right)^2 \, . \qquad (4.14)$$

Furthermore, for the integrations involved in equations (4.13-4.14), $r$ and $h$ are now in units of $GM/c^2$ ($G$ is the gravitational constant, $M$ is the central object mass, and $c$ is the speed of light) and $\mathcal{F}$ can be considered completely dimensionless, having all dimensional constants factored out from the top and bottom integrals.

As an example, let us look at the case of an optically thick, gas pressure dominated disc. For this case, the photon flux is a function of the effective temperature, which in turn is a simple function of the radius. We find

$$\mathcal{F}^2 \propto T_{eff}^6 \propto r^{-9/2} \sim 6^{-9/2} \left(\frac{f_L}{f_i}\right)^3 \, . \qquad (4.15)$$

For the disc thickness we have

$$h \sim h_0 \left(\frac{f_L}{f_i}\right)^{-7/10} \propto r^{21/20} \, , \qquad (4.16)$$

where

$$h_0 \sim 0.1 \, \mathcal{L}^{1/5} \alpha^{-1/10} \mathcal{M}^{-1/10} \, , \qquad (4.17)$$

and $\mathcal{L}$, as before, is the disc luminosity expressed as a fraction of the Eddington luminosity and $\mathcal{M}$ is the central object mass in units of $M_\odot$ (*cf.* Shakura & Sunyaev 1973). Thus we find

$$\mathcal{P}(f) \sim 3 \times 10^{-4} \, f_i^{-1} \, \frac{\mathcal{L}^{1/5} \alpha^{19/10}}{\mathcal{M}^{1/10}} \left(\frac{f}{f_i}\right)^{-7/10} \, , \qquad (4.18)$$

for $f < f_i$. The frequency on the inner edge of the disc is $f_i \sim 2.2 \times 10^3/\mathcal{M}$ Hz. This is well beyond the range of most detectors, so we choose to normalize the PSD to a frequency of $10^2$ Hz. We then have

$$\mathcal{P}(f) \sim 1 \times 10^{-7} \left(\frac{\mathcal{L}}{0.1}\right)^{1/5} \left(\frac{\mathcal{M}}{6}\right)^{1/5} \left(\frac{\alpha}{0.3}\right)^{19/10} \left(\frac{f}{10^2 \text{ Hz}}\right)^{-7/10} \, , \qquad (4.19)$$

for $f < f_i$. (Note that more realistic flux laws rollover near the inner edge of the disc and hence produce fewer photons in the inner regions than implied by eq. [4.15]. This will tend to decrease $\mathcal{N}$ and hence increase the amplitude of th PSD by a factor of several.)

The slope and amplitude of the PSD are in very good agreement with the data obtained from GS 1124-683 as presented in Figure 1. The data in this figure are taken from Miyamoto et al. (1993) and show the $\mathcal{P}(f) \propto f^{-0.7}$ characteristic of black hole high states. Note that for $f < f_i$ the theoretical PSD slope is entirely determined by the radial dependence of the disc thickness, whereas for $f > f_i$ the slope is determined by the physics of the turbulent cascade. Specifically, if we assume $h \propto r^\beta$ and that for the turbulent cascade the turbulent kinetic energy per wavenumber, $E(k)$, scales with wavenumber, $k \equiv \ell^{-1}$, as $E(k) \propto k^{-\nu}$, we obtain

$$\begin{aligned} \mathcal{P}(f) &\propto \frac{\alpha^2 h_0}{f_i} \left(\frac{f}{f_i}\right)^{-2\beta/3} \qquad \left[\left(\frac{f}{f_i}\right) \leq 1\right] \\ &\propto \frac{\alpha^2 h_0}{f_i} \left(\frac{f}{f_i}\right)^{-(5+\nu)/(3-\nu)} \qquad \left[\left(\frac{f}{f_i}\right) > 1\right] \; . \end{aligned} \tag{4.20}$$

For the Kolmogorov cascade $\nu = 5/3$, and therefore the high frequency power rolls over as $f^{-5}$.

That the slope and amplitude from equation (4.19) match the observations so well is most likely merely fotuitous, as we have not included the detector bandpass. Equation (4.19) counts photons of *all* energies, whereas the data of Figure 1 is for the bandpass $1.2 - 15.7$ keV. Including this bandpass will tend to exclude the low Keplerian frequency outer regions of the disc, which emit predominantly low energy photons. Below, we explore two accretion disc models for which we compute the effects of including the bandpass.

The first model is a variant of the Shakura-Sunyaev $\alpha$-disc model (Nowak 1992). The disc is in Keplerian rotation starting from the marginally stable orbit at $r = 6$ out to infinity (throughout, we shall take the radius, $r$, and disc half-thickness, $h$, to be in units of $GM/c^2$). The flux from a given radius is derived by assuming energy conservation as mass flows from one Keplerian rotating ring to another. The energy flux from both sides of the disc is then given by

$$F(r) = 1.24 \times 10^{28} \text{ erg cm}^{-2} \text{ s}^{-1} \; \frac{\mathcal{L}}{\mathcal{M}} \mathcal{I} \, t \, r^{-3} \; , \tag{4.21}$$

where $\mathcal{L}$ and $\mathcal{M}$ are defined as before,

$$s \equiv \left(1 - \frac{6}{r} + \frac{36}{r^2}\right) \; , \quad t \equiv \left(1 - \frac{8}{r} + \frac{60}{r^2}\right) \; , \tag{4.22a}$$

and the function

$$\mathcal{I} \equiv \left(1 - \sqrt{\frac{6}{s \, r}}\right) \; , \tag{4.22b}$$

arises as a consequence of assuming that the viscous torque vanishes at the disc inner edge (*cf.* Shakura & Sunyaev 1973). The disc is broken into radiation pressure dominated and gas pressure dominated regions. (For the most part we will ignore the issue of disc instabilities, though see the discussion of §5.) In general, the radiation pressure dominated regions are located in the inner regions of the disc, and the width increases as the disc luminosity increases. In the radiation pressure dominated regions we take the disc half-thickness (in units of $GM/c^2$) to be

$$h \approx 18. \, \mathcal{L} \, \mathcal{I} \; , \tag{4.23}$$

while in the gas pressure dominated regions we take

$$h \approx 1.7 \times 10^{-2} \left(\frac{\mathcal{L}^2}{\alpha \mathcal{M}}\right)^{1/10} \mathcal{I}^{1/5} \, r^{21/20} \quad . \tag{4.24}$$

(The transitions between the regions are easily calculated from the disc structure equations; *cf.* Shakura & Sunyaev 1973, Shapiro & Teukolsky 1983.)

Since we are assuming that the disc is optically thick throughout, the emitted spectrum will be either a blackbody spectrum, or a modified blackbody spectrum (*cf.* Shapiro & Teukolsky 1983). For the blackbody spectrum, we assume that at each radius the disc emits a blackbody spectrum at a temperature $T_{eff}$ given by

$$kT_{eff} = 10.5 \text{ keV} \left(\frac{\mathcal{L}}{\mathcal{M}}\right)^{1/4} t^{1/4} \, r^{-3/4} \, \mathcal{I}^{1/4} \quad . \tag{4.25}$$

Note that this temperature peaks at a value $kT_{eff} \sim 1$ keV $(\mathcal{L}/\mathcal{M})^{1/4}$, which is typically a lower value than is observed for the peak temperature of black hole candidate systems in their high state. In the electron scattering opacity dominated regions of the disc, we expect the emitted spectrum to take on a modified blackbody form, for which the flux per unit frequency is given by

$$F_\nu \propto \frac{z^{3/2} \exp(-z/2)}{[\exp(z) - 1]^{1/2}} \quad , \tag{4.26}$$

where $z \equiv h\nu/kT_s$, and $T_s$ is the modified blackbody temperature (*cf.* Shapiro & Teukolsky 1983). For the gas pressure dominated regions of the disc this temperature is given by

$$kT_s \approx 2.5 \times 10^3 \text{ keV} \left(\frac{\alpha^2 \mathcal{L}^8}{\mathcal{M}^2}\right)^{1/9} \mathcal{I}^{8/9} \, r^{-5/3} \quad , \tag{4.27}$$

while for the radiation pressure dominated regions of the disc we take

$$kT_s \approx 24 \text{ keV} \left(\frac{\alpha^7 \mathcal{L}^{16}}{\mathcal{M}^{13}}\right)^{1/45} \mathcal{I}^{16/45} \, r^{-87/90} \quad . \tag{4.28}$$

In practice, we use a modified blackbody spectrum whenever $kT_{eff} < 2.25 \, kT_s$. (At this transition point, the bolometric *photon* fluxes of the blackbody and modified blackbody spectra are identical.)

Using the disc structure equations, we self-consistently calculate the transitions between radiation and gas pressure dominance, and blackbody and modifed blackbody spectra. The positions of these transitions, as well as the value of the various disc temperatures, are dependent upon $\alpha$, $\mathcal{L}$, and $\mathcal{M}$. Many of the black hole candidates with well determined mass functions appear to have masses near $M \sim 6 \, M_\odot$, therefore we shall set $\mathcal{M} = 6$. Furthermore, for the high, quiet state of black hole candidates, $\mathcal{L} \sim 0.1$ (*cf.* Nowak 1994b); therefore, we shall explore the parameters $\mathcal{L} = 0.03, 0.1, 0.3$. At each of these luminosities, we calculate a PSD for the parameters $\alpha = 0.03, 0.1, 0.3, 1.0$. We use the same bandpass as was used for the GS 1124-683 data presented in Figure 1, namely 1.2 − 15.7 keV. The bandpass is included in equations (4.13) and (4.14) by multiplying the photon number flux, $\mathcal{F}(r)$, by the fraction of photons emitted into the bandpass at that radius. Results for these 12 runs are presented in Figure 1.

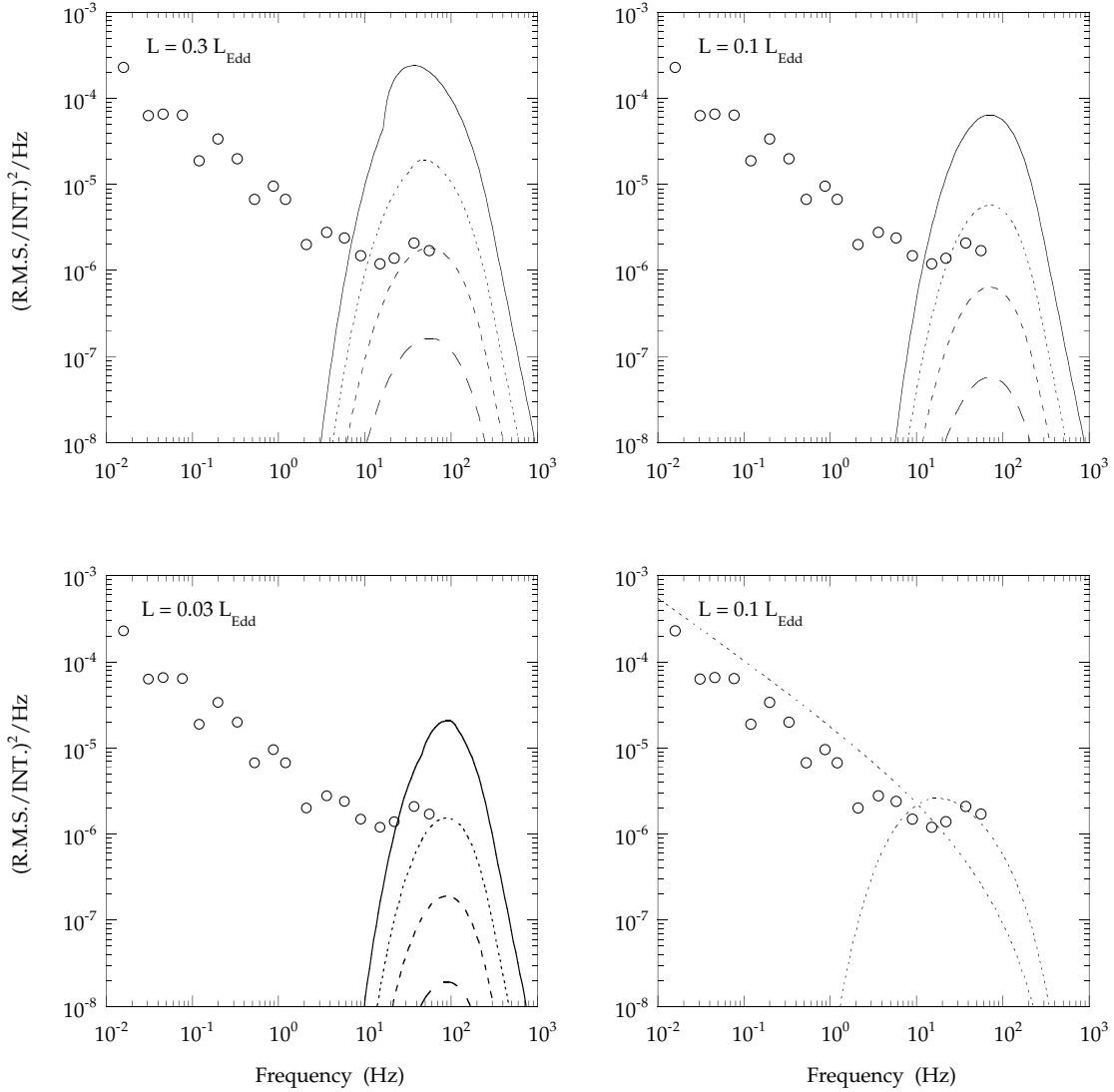

**Figure 1.** Comparison of theoretical acoustic mode PSD with GS 1124-683 data (circles; *cf.* Miyamoto et al. 1993). In all graphs, solid line: $\alpha = 1$, dotted line: $\alpha = 0.3$, short dash: $\alpha = 0.1$, long dash: $\alpha = 0.03$. The disk models were as follows. Upper left: Shakura & Sunyaev disk with 30% of the Eddington luminosity. Upper right: Shakura & Sunyaev disk with 10% of the Eddington luminosity. Lower left: Shakura & Sunyaev disk with 3% of the Eddington luminosity. Lower right: Blackbody at each radius with $T^4$ proportional to the local flux and the peak temperature artificially set to 1 keV at 10% Eddington luminosity. For all models, the central object mass was 6 $M_\odot$, and the bandpass was $1.2 - 15.7$ keV (except for the long dotted line of the lower right figure which included *all* photons).

We compare these results to the PSD of GS 1124-683 in its high, quiet state. For these observations, $\mathcal{L} \sim 0.1$ (Miyamoto et al. 1993). There are several things to note here. First, as expected, including the bandpass drastically reduces the low frequency power. The power is reduced to such an extent that none of the models have as much power as the observations below $f \sim 10$ Hz. Second, we require a fairly large $\alpha$ in order to fit the high frequency observations.

This is because larger $\alpha$ implies both larger acoustic mode amplitudes (*cf.* equation [4.11]) as well as higher disc temperatures that give significant flux in our bandpass (*cf.* equations [4.27], [4.28]). Setting $\alpha = 1.0$ clearly gives too much power, $\alpha = 0.3$ is in reasonable agreement with the high frequency observations, and $\alpha < 0.3$ for the most part gives too little power. Unfortunately, the observations are below that of the expected rollover in the acoustic mode power. If the high frequency power is really due to disc turbulence, then high frequency power spectra – such as should be available with the upcoming experiments on the XTE and USA satellites – should show a sharp drop in the PSD above $f \sim 10^2$ Hz.

Since the emitted spectrum of an accretion disc is highly uncertain, we explore one other accretion disc model. For this model we choose the energy flux to have the same functional dependence as in equation (4.21); however, we assume that the disc emits as a blackbody throughout but artificially set the peak temperature to 1 keV. We choose 1 keV as being typical of the observations, and take $T_{eff} \propto F^{1/4}$ as being the least parameter dependent approximation (there is no dependence upon the unknown parameter $\alpha$, and by artificially setting the peak temperature to 1 keV we have removed the dependence upon $\mathcal{L}$ and $\mathcal{M}$). We calculate the PSD for this model both with the 1.2 − 15.7 keV bandpass and without it. For these calculations we set $\mathcal{L} = 0.1$, $\mathcal{M} = 6$, and $\alpha = 0.3$. All three parameters determine the radii at which the disc makes a transition between radiation pressure dominance and gas pressure dominance, and furthermore $\alpha$ determines the amplitude of the acoustic modes. The results are also presented in Figure 1. Note that the simulation with no bandpass agrees with the data quite well; however, adding in the bandpass cuts off the low frequency power. There is reasonable agreement between the simulation and the data for $f \sim 3 - 10^2$ Hz, which is a slightly broader region of agreement than the Shakura-Sunyaev disc. However, there is still two and a half orders of magnitude of low frequency data that cannot be explained by acoustic modes.

## 5 VISCOUS/THERMAL EFFECTS

As we have seen from the previous section, fluctuations on turbulent (dynamical) time scales can only account for the observed power between a few and several hundred Hz. Fluctuations on time scales ranging from $\sim 10^{-2} - 3$ Hz must arise from other sources. The purpose of this section is to show that a plausible source of this low frequency power is weak variations of the disc flux on local viscous/thermal time scales. We expect that such fluctuations might arise from disc instabilities (*cf.* Shakura & Sunyaev 1976, Piran 1978). The standard Shakura-Sunyaev $\alpha$-disc becomes unstable to thermal and viscous instabilities at $\mathcal{L} \sim 0.03$ for typical parameters. If $\mathcal{L} \gtrsim 0.1$, nearly the entire X-ray emitting region of the disc is subject to viscous and thermal instabilities. Below we will not consider any detailed mechanisms for disc instabilities; rather, we will merely show that weak flux variations on characteristic instability timescales can plausibly account for the low frequency power.

We assume that locally the accretion rate is only steady when averaged over times greater than the local viscous/thermal time scales. For times shorter than this we assume that the flux varies exponentially in time with some fractional maximum amplitude, $\jmath$, over a radial extent, $\Delta r$. The variation in the photon count rate, $\Delta x$, emitted from radius $r$ then becomes

$$\Delta x \ = \ 2\pi \ r \Delta r \ \jmath \mathcal{F} \ . \qquad (5.1)$$

We know that integrating the power spectral density over frequency yields a quantity proportional to the square of the above variation. Furthermore, exponential variations yield PSDs that have functional forms $\mathcal{P}(f) \propto 1/[1 + (2\pi f \tau)^2]$, where $\tau$ is the viscous time scale. We therefore adopt this as the functional form for the PSD at a given radius (*cf.* Nowak 1994a). We find the amplitude, $\mathcal{A}$, of the PSD of this radius by setting

$$\Delta x^2 = \int_0^\infty \frac{\mathcal{A}}{1 + (2\pi f \tau)^2} \, df \; , \tag{5.2}$$

which yields

$$\mathcal{A} = 16\pi^2 \, \tau \, r^2 \Delta r^2 \, j^2 \mathcal{F}^2 \; . \tag{5.3}$$

The power spectral density of the disc as a whole then becomes an incoherent sum of the PSDs of the individual disc radii, normalized by the square of the total luminosity of the disc. As an example, we take the instability time scales over which the flux varies to range from $\tau = [\alpha(h/\lambda)^2 \Omega]^{-1}$ to $\tau = (\alpha\Omega)^{-1}$, where as before $\alpha$ is the viscosity parameter, $\Omega$ is the Keplerian frequency, and now the wavelength of the instability is given by $\lambda$ ($h \lesssim \lambda \ll r$). (Note that for short wavelengths classical viscous and thermal instabilities merge into a single branch; Shakura & Sunyaev 1976.) As a simple example, let us take $\Delta r \sim \lambda \propto \mathcal{C} r$, where for our purposes $\mathcal{C}$ will be a dimensionless constant. For example, we might imagine that in a gas pressure dominated disk $\lambda \sim h \sim 0.02 \, \mathcal{L}^{1/5} \, r^{21/20}$, giving $\mathcal{C} \sim 10^{-2}$ and nearly constant over a large range of radii. For this case we would be saying that the accretion rate is unsteady on radial scales less than a disc scale height, or equivalently the largest turbulent eddy size. We now write for the PSD of the disc:

$$\mathcal{P}(f) \sim \frac{\sum \frac{16\pi^2}{[1 + (2\pi f/\alpha'\Omega)^2]} (\alpha'\Omega)^{-1} \, r^2 \, \lambda^2 \, j^2 \, \mathcal{F}^2}{\left(\int dr \, 2\pi r \, \mathcal{F}\right)^2} \; , \tag{5.4}$$

where the sum is over radial bins of width $\Delta r \sim \lambda$ and $\alpha'$ is the constant of proportionality between the inverse instability time scale and the Keplerian frequency (equal to $\sim \alpha$ for thermal instabilities and/or short wavelengths and equal to $\sim [h/\lambda]^2 \alpha$ for viscous instabilities).

We analytically explore the behavior of this solution by substituting in the simple flux law $\mathcal{F} \propto r^{-9/4}$, along with the Keplerian frequency $\Omega = c^3/GM \, r^{-3/2}$. Also, we replace the sum by a logarithmic integral over $dr/\mathcal{C}r$. This yields

$$\mathcal{P}(f) \approx \frac{\sqrt{6}}{4} \frac{\mathcal{C} j^2}{\alpha'} \frac{GM}{c^3} \int_6^\infty \frac{1}{1 + a^2 r^3} \, dr \; , \tag{5.5}$$

where $r$ is now in the units of $GM/c^2$ and $a \equiv (2\pi f GM/\alpha' c^3)$. The above integral can be performed analytically; however, it is a bit more enlightening to look at the limits $a \ll 1$ and $a \gg 1$. In these limits, we obtain

$$\begin{aligned} \mathcal{P}(f) \approx & \; \frac{\sqrt{2}}{6} \pi \frac{\mathcal{C} j^2}{\alpha'} \frac{GM}{c^3} \, a^{-2/3} & (a \ll 1) \\ & \frac{1}{576\sqrt{6}} \frac{\mathcal{C} j^2}{\alpha'} \frac{GM}{c^3} \, a^{-2} & (a \gg 1) \; . \end{aligned} \tag{5.6}$$

The low frequency PSD has a slope that agrees well with observations (though we note that if we had chosen $\Delta r \propto \mathcal{C}$ we would have found $\mathcal{P}(f) \propto \ln[1 + a^{-2}/216]$ instead). If we set $M = 6\ M_\odot$ we then have

$$\mathcal{P}(f) \approx \begin{array}{ll} 0.15\ \dfrac{\mathcal{C}\ \jmath^2}{\alpha'^{\ 1/3}} \left(\dfrac{f}{10^{-2}\ \text{Hz}}\right)^{-2/3} & (a \ll 1) \\[1em] 6 \times 10^{-5}\ \mathcal{C}\ \jmath^2 \alpha' \left(\dfrac{f}{10^2\ \text{Hz}}\right)^{-2} & (a \gg 1) \end{array} \quad (5.7)$$

The low frequency solution has an amplitude in reasonable agreement with the observations for $\mathcal{C} \sim 0.01$, $\alpha' \sim 0.1$, and $\jmath \sim 0.3$. That is, flux variations on the order of 30% occuring over extents of roughly a scale height yield power comparable to the observations. Note, however, that with the same parameters the high frequency solution yields too little power.

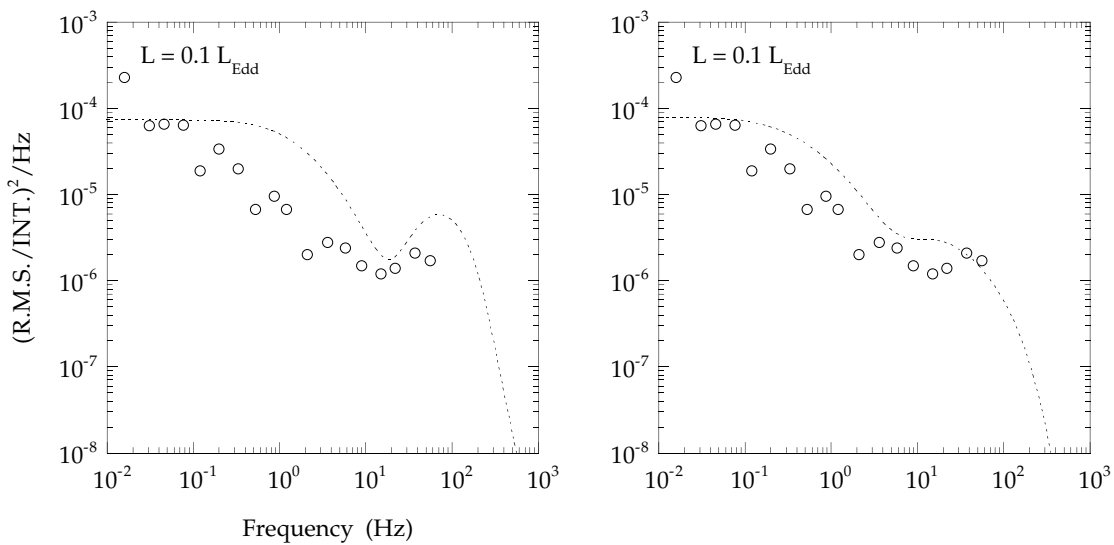

**Figure 2.** Combination of viscous/thermal time scale and acoustic mode power compared to the observed PSD of GS 1124-683 (circles; *cf.* Miyamoto et al. 1993). In both graphs we have set $M = 6\ M_\odot$, $\mathcal{L} = 0.1$, $\alpha = 0.3$, and $\alpha' = 0.03$. As in Figure 1, the bandpass for both graphs was taken to be $1.2 - 15.7$ keV. Left: Shakura & Sunyaev disk (*cf.* equations [4.21]-[4.28]) with $\mathcal{C}\jmath^2 = 1.3 \times 10^{-3}$. Right: Flux is blackbody at each radius (*cf.* eq. [4.21]) with the peak temperature being artificially set to 1 keV and $\mathcal{C}\jmath^2 = 8 \times 10^{-4}$.

In the above, we have used an oversimplified flux model and have not included a bandpass. As before, a bandpass in the soft X-ray range will exclude the outer regions of the disc, and hence lead to a reduction of the low frequency power. In addition, for accretion discs with "no-torque" inner edge boundary conditions the flux drops to zero on the inner edge of the disc. This will lead to a further reduction in the high frequency power. In Figure 2, we present numerical calculations wherein we use a bandpass of $1.2 - 15.7$ keV and use the flux laws and disc models of §4. The left graph is for a Shakura-Sunyaev disc (*cf.* equations [4.21]-[4.28]) with $M = 6\ M_\odot$, $\mathcal{L} = 0.1$, and $\alpha = 0.3$. We have also taken $\alpha' = 0.03$ (which for a viscous instability would correspond to $h/\lambda \sim 0.3$) and we have set $\mathcal{C}\jmath^2 = 1.3 \times 10^{-3}$ (*i.e.* $\sim 30\%$ flux variations over radial extents the order of the disk thickness). The right graph is for the flux law of equation (4.21) (*i.e.* blackbody), with the peak temperature again being artificially set to 1 keV. Here also we have taken $M = 6\ M_\odot$, $\mathcal{L} = 0.1$, $\alpha = 0.3$, and $\alpha' = 0.03$. We have set $\mathcal{C}\jmath^2 = 8 \times 10^{-4}$. In each of these graphs we have added in the dynamical power as presented in Figure 1 (for the same parameters). As before, the circles represent high state data of GS 1124-68, for which $M \sim 6\ M_\odot$ and $\mathcal{L} \sim 0.1$. (Note that in this figure we have not attempted to find a rigorous "best fit" to the data. Our parameters were chosen by eye in order to give a general indication of the trends.)

The combination of viscous and dynamical power is mildly successful in reproducing the data. The amplitude and rough shape of the low and high frequency data are reproduced, however, there are significant deviations. Note that the shape of the low frequency power for the Shakura-Sunyaev discs tends to be flat at very low frequencies, and then rolls over into the $f^{-2}$ behavior. The expected $f^{-2/3}$ behavior is absent. This is because the high frequency cut-off (due to the no-torque boundary condition) and the low-frequency cut-off (due to the bandpass) restricts us to a narrow range of radii. We are mostly seeing the $\mathcal{P}(f) \propto 1/[1 + (2\pi f\tau)^2]$ behavior from this narrow range of radii. The blackbody flux law of equation (4.21) with the peak temperature set to 1 keV is observable over a slightly broader range of radii, so we see some of the expected $f^{-2/3}$ behavior, which is in better agreement with the data. This model also has a broader high frequency peak, which also agrees better with the data. We see that in order to reproduce the data with viscous/thermal and dynamical time scales, we require a disc that emits soft X-rays over a broad range of radii. It is unlikely, however, that any realistic flux model will allow for much power above $\sim 10^2$ Hz. Furthermore, we expect that with most flux models the PSD must also flatten below $\sim 10^{-2}$ Hz. Again, observations by XTE and USA will be able to confirm or refute this behavior.

## 6 CONCLUSIONS

If we are to look for observational clues that elucidate the basic physics of accretion discs we should explore the high state of galactic black hole candidates. The high state appears to be well explained by optically thick accretion disc models and appears to be relatively quiet and stable on short time scales. This is to be compared to the non-thermal and highly variable low state, which may require the addition of clouds, winds, and/or optically thin regimes in order to explain the observed energy spectra. In this work we have explored the consequences of the assumption that viscosity (and hence angular momentum transport and energy generation) is ultimately due to accretion disc turbulence.

We explored two ways in which disc turbulence could modulate the observed emission. First, we considered the role of advection; however, we found that typical photon diffusion times through a scattering atmosphere are shorter than eddy rollover times. Thus eddies are incapable of "dredging" hot photons from the disc midplane. Second, we considered direct modulation of the emission rate. Under the assumption that locally within the disc the energy generation rate scales as a small power of the density, one expects flux modulations to arise from the weak density fluctuations associated with the turbulence. These weak density fluctuations can be considered to be essentially acoustic modes driven by the turbulence. The amplitudes of these acoustic modes are proportional to the Mach number of the turbulence, which in our model is related to the familiar disc parameter $\alpha$. Note that our model assumed isotropic hydrodyamic turbulence, whereas many recent models employ either two dimensional hydrodynamic turbulence or three dimensional magnetohydrodynamic turbulence (*cf.* the references of §1). For our model, *at least* the largest eddies (with the lowest frequencies) were capable of modulating the flux.

We then considered how this variability would be manifested in the observed X-ray power spectral density (PSD). We found that the theoretical PSD could be derived from summing the flux variations of individual modes over the mode phase space density. We compared these predictions to high state observations of GS 1124-683, which show the characteristic low amplitude $\mathcal{P}(f) \propto f^{-0.7}$ of BHC high states (*cf.* Miyamoto 1994). Our model was able to reproduce much of the observed high frequency power; however, when we included the X-ray bandpass of the detectors we were unable to reproduce the low frequency power. In addition, we required a large $\alpha$ ($\gtrsim 0.3$) in order to reproduce the amplitude of the high frequency power. This large an $\alpha$ is inconsistent with estimates of two dimensional hydrodynamic turbulence (*cf.* Narayan et al. 1994); however, it is as yet unclear whether or not this large an $\alpha$ is inconsistent with MHD turbulence as embodied in the magnetic shearing instability framework (*cf.* the references of §1). Note also that in equation (4.13) we have assumed that eddies of *all* sizes are capable of modulating the emission. We have not accounted for the fact that at some radii and depths only the largest eddies contribute to the modulation. Taking this detail into account would further reduce the high frequency power; however, this should not drastically effect the simple estimates presented in §4.

Finally, we considered the possibility that the low frequency power could be caused by weak fluctuations on viscous/thermal time scales. We saw that flux variations on the order of 30% over distances comparable to the largest eddy sizes could account for the amplitude and rough shape of the low frequency power. However, viscous/thermal fluctuations were incapable of simultaneously reproducing the low and high frequency power. Combining the acoustic mode power with these crude fluctuation estimates was roughly capable of reproducing the observed PSD over three-four decades of frequency, though there were noticeable deviations.

We note that it is difficult for any single variability mechanism to reproduce four decades of power *if* the variabilty time scale is proportional to the Keplerian time scale. The X-ray producing range of the disc exists between radii of $r \approx 6 - 150 \ GM/c^2$. Since the Keplerian time scale goes as $r^{3/2}$, any variability mechanism proportional to this time scale can cover only two orders of magnitude of frequency in the X-ray producing region of the disc. We see that two separate time scales can cover perhaps four decades, as we showed in §5; however, unless there is a third mechanism at work we expect there to be a break for $f \lesssim 10^{-2}$ Hz. In addition, as the dynamical time scale is likely to be the fastest time scale in the disc we expect a sharp rollover for $f \gtrsim 10^2$ Hz. For our model of turbulence, this rollover should be $\propto f^{-5}$. We expect that both XTE and USA should be able to observe the low frequency break and high frequency rollover if our ideas are correct.

We see two major areas in which the above model can be improved: the model of turbulence, and the model of emission from the disc. We expect that any significant progress made will come from attacking both of these problems simultaneously. Of great interest would be a more complete understanding (in a global sense) of the magnetic shearing instability, if this is indeed the correct explanation for accretion disc viscosity. One needs to understand the amplitude of the turbulence, how it couples to the energy generation rate, and the characteristic viscous time scale. Also, as we have seen from Figure 2, the particular flux law that we use affects the range of frequencies over which the the emission is modulated. It is unlikely, however, that any reasonable flux law will provide for dynamical variability power at frequencies much below a few Hz unless some mechanism such as Comptonization is invoked. We note though that the high state lacks an extended, hard tail in the energy spectra, and therefore it is difficult to reconcile such mechanisms with the observations.

Finally, we need to ask whether such a thing as accretion disk turbulence exists at all in these black hole candidate systems. To this end we note that the $\mathcal{P}(f) \propto f^{-0.7}$ dependence appears to be a characteristic behavior of the thermal emission of BHC. Even when model fits indicate that less than 1% of the photon flux is emanating from a decidely non-thermal source, the $\mathcal{P}(f) \propto f^{-0.7}$ dependence remains – albeit at a very low amplitude (Miyamoto 1994). As the slope of the PSD implies, the integrated power must peak near the characteristic dynamical time scales of the disc. The exact nature of these characteristic oscillations are still in doubt; however, these weak variations seem to be fundamental to the emission mechanisms of this BHC state and perhaps are the best evidence for the existence of accretion disc turbulence.

## ACKNOWLEDGEMENTS

The authors would like to acknowledge useful conversations with Lars Bildsten, Norm Murray, Chris Thompson, and Brian Vaughan. We would also like to thank James Chiang and Derek Richardson for graphical support.